\documentclass[12pt,preprint]{aastex}

\def\phil{\hbox{$\phi^{\rm i}_{\rm l}$}}
\def\phit{\hbox{$\phi^{\rm i}_{\rm t}$}}
\def\gammai{\hbox{$\Gamma^{\rm i}$}}
\def\etai{\hbox{$\eta^{\rm i}$}}
\def\remi{\hbox{$r^{\rm i}_{\rm em}$}}
\def\sii{\hbox{$s^{\rm i}_{\rm L}$}}
\def\rlc{\hbox{$r_{\rm LC}$}}
\def\degsp{\hbox{$^{\circ\/\ }$}}

\shorttitle{Emission geometry of radio pulsars}
\shortauthors{Gupta and Gangadhara}
\received{}

\begin{document}
\title{Understanding the radio emission geometry of multi-component radio pulsars from retardation 
and aberration effects}

\author{Y. Gupta\altaffilmark{1} and R. T. Gangadhara\altaffilmark{2}}

\affil{\altaffilmark{1}National Astronomy and Ionosphere Center, Arecibo Observatory, HC3 Box 53995, Arecibo, PR 00612, USA}
\affil{\altaffilmark{2}Indian Institute of Astrophysics, Bangalore 560034, India}

\altaffiltext{1}{Permanent Address: NCRA (TIFR), Pune University Campus, Pune 411007, India;  E-mail: {\tt ygupta@ncra.tifr.res.in }}
\altaffiltext{2}{E-mail: {\tt ganga@iiap.ernet.in}}

\begin{abstract}
We have conducted a detailed analysis of the emission geometry of a 
handful of radio pulsars that have prominent, multiple-component profiles
at meter wavelengths.  From careful determination of the total number of 
emission components and their locations in pulse longitude, we find 
that all of the six pulsars show clear evidence for retardation and 
aberration effects in the conal emission beams.  Using this information, 
coupled with a dipolar field geometry, we obtain estimates of the height 
and transverse location in the magnetosphere, for each of the emitting 
cones in these pulsars.  These results support our earlier conclusions 
for PSR~B0329+54 in that we find successive outer cones (in cases of 
multi-cone pulsars) being emitted at higher altitudes in the magnetosphere.  
The range of inferred heights is from $\sim$ 200 to $\sim$ 2200 km.  The 
set of ``active'' field lines from which the conal emissions originate are 
located in the region from $\sim$ 0.22 to $\sim$ 0.74 of the polar cap radius. 
At the neutron star surface, these conal rings map to radii of a few to 
several tens of meters and the separation between successive rings is about 
10 to 20 meters.  We discuss the implications of these findings for the 
understanding of the pulsar emission geometry and for current theories and 
models of the emission mechanism. 
\end{abstract}

\section{Introduction }
Radio emission from a pulsar is believed to originate in the open 
field line region of the polar cap of the neutron star.  The size, 
shape and location of regions of radio emission in the average 
profiles of pulsars is thus expected to reflect the arrangement of 
emission regions in the pulsar magnetosphere.
Pulsar average profiles exhibit a great diversity in shape, and their
classification based on the number of emission components is a useful 
starting point to study the emission characteristics of pulsars. 
Rankin (1993, 1990; and references therein) has carried out 
such a detailed classification, and has come up with the conclusion 
that there are two kinds of emission components -- core and conal -- 
in pulsar profiles, which result from two distinct types of emission 
mechanisms.  Further, Rankin proposes that the conal components arise 
from two nested hollow cones of emission, which along with a central 
core emission region, make up the complete pulsar emission beam.  The 
actual profile observed for a given pulsar depends on the cut that 
the observer's line-of-sight makes through this emission beam.  From 
the above work, Rankin also concludes that core radiation originates 
from very close to the neutron star surface whereas the conal radiation 
comes from regions higher up in the magnetosphere.  The outer cone is 
postulated to originate higher up in the magnetosphere than the inner 
cone, but along the same set of field lines.
Recent work by \cite{GGGK} which compares frequency dependence of
pulsar radiation features for average profiles with those for single 
pulses, comes up with evidence in support of such a conal beam model.
A somewhat different model of the pulsar emission beam is described by 
\cite{LM88} who propose that the emission within the beam is patchy, 
i.e. the distribution of component locations within 
the beam is random, rather than organized in one or more hollow cones. 

One of the problems in resolving this conflict is related to a correct determination
of the total number of emission components for a pulsar and the significance of
their arrangement within the pulse window.
In a recent paper (Gangadhara \& Gupta, 2001; hereafter paper I), we have addressed
these issues with the help of a novel method of studying the emission geometry of 
pulsars.  Using the sensitive ``window-thresholding'' technique (W-T technique hereafter) 
on single pulse data for a given pulsar, we first identify all possible conal emission 
components in the profile that can be detected given the sensitivity of the data, 
and determine accurate locations in pulse longitude for these.  
From this we determine the total number and arrangement of emission cones in the 
observed profile.  Then, from detection of retardation and aberration effects in the 
locations of the conal components, we estimate the emission height for each cone.  
Coupling this with a dipolar field geometry and the available information about the 
relative orientations of the spin and rotation axes of the pulsar and the line of sight, 
we are able to identify the field lines from which each cone originates, thereby giving 
a complete solution for the emission geometry for each cone.  Using this technique on 
single pulse data for PSR~B0329+54 at two different radio frequencies, we showed in 
paper I that there are as many as 9 identifiable components in this pulsar, arranged 
in the form of 4 concentric rings around the central core component.  The emission 
heights for the cones were found to increase systematically from inner to outer cones 
for a given frequency, ranging from $\sim$ 160 to $\sim$ 1150 km.  Also, for a given 
cone, the height was found to reduce with increasing frequency, providing direct proof 
of the commonly accepted ``radius to frequency mapping'' paradigm.

In this paper, we extend our study of emission geometry to another six pulsars.
Section 2 describes the observations and data analysis steps.  In section 3, we present 
our results, where we are able to solve for the emission geometry for 5 of the pulsars 
studied, and have tentative results for the sixth.
In section 4 we discuss the implications of our findings for the current understanding 
of pulsar emission physics.

\section{Observations and Data analysis}

\begin{table}
\begin{center}
\caption{Properties of the pulsars used for emission geometry studies. \label{tbl1}}
\begin{tabular}{cclccccc}
\tableline \tableline
Pulsar & Frequency & Period  & $\alpha$     & $\beta$      & profile & \# of single  & \# of new \\
       & \rm (MHz) & \rm (s) & \rm (deg) & \rm (deg) & class   & pulses  & components \\
\tableline
\noalign{\smallskip}
PSR~B0450$-$18 & \rm 318 & \rm 0.5489   & \rm 24.0 &\rm 4.0 & T & 2405  & ? \\
PSR~B1237+25   & \rm 318 & \rm 1.3824   & \rm 53.0 &\rm 0.0 & M & 1915  & 2 \\
PSR~B1821+05   & \rm 318 & \rm 0.7529   & \rm 32.0 &\rm 1.7 & T & 1435  & 4 \\
PSR~B1857$-$26 & \rm 318 & \rm 0.6122   & \rm 25.0 &\rm 2.2 & M & 2150  & 2 \\
PSR~B2045$-$16 & \rm 328 & \rm 1.9617   & \rm 36.0 &\rm 1.1 & T & 1680  & 3 \\
PSR~B2111+46   & \rm 333 & \rm 1.0147   & \rm 9.0  &\rm 1.4 & T & 2900  & 4 \\
\tableline
\end{tabular}
\end{center}
\end{table}

For our study of emission geometry, we selected a set of pulsars with clearly 
defined properties that make it easy to apply the techniques used 
for PSR~B0329+54 in paper I.  These included the following criteria: 
(i) the presence of multiple emission components (or indications thereof) 
in the existing profiles; (ii) the presence of a reasonably well identified core 
component at our frequency of observation, 
which was the 325 MHz band of the GMRT; (iii) strong enough mean flux such that
reasonable signal to noise ratio (SNR) could be expected for single pulse observations 
with the incoherent array mode of the GMRT.  The first two constraints above confine 
the selected pulsars to the category of ``triple'' or ``multiple'' pulsars as defined 
by \cite{JMRd}, or equivalently, to the category of ``cones with cores'' pulsars, 
as defined by \cite{LM88}.  
Of the possible candidates, we avoided pulsars with very wide profiles (e.g. 
PSR~B0826$-$34, PSR~B1541+09) as it is not easy to unambiguously identify discrete
emission components in these profiles due to the extended emission over a large fraction
of the pulse period.  The final list of pulsars selected for our study is given in 
Table~\ref{tbl1}, along with relevant details.  Here, $\alpha$ and $\beta$ refer to
the standard definitions of the angle between the rotation and magnetic axes and
the impact angle of the line of sight with respect to the magnetic axis, respectively.  
The values for these angles are taken from \cite{JMRd}. These are not very different 
from the corresponding values given by \cite{LM88} -- the typical difference for $\alpha$ 
is $\sim$ a few degrees and that for $\beta$ is less than $\sim$ 0.5 to 1.0 degree.

The observations were carried out at the GMRT, in the 325 MHz band, using the incoherent
array mode (see \cite{Gupta} for more details about the pulsar modes of operation of the 
GMRT). The data were obtained by incoherent addition of the dual polarization signals 
from 12 antennas.  The bandwidth used was 16~MHz, divided into 256 spectral channels by 
the digital back-ends.  The raw data were integrated to a time resolution of 0.516 
milliseconds before being recorded for off-line analysis, where the data were dedispersed 
and gated to obtain the single pulse sequence for each pulsar.  During this analysis, care 
was taken to check and flag the data for radio frequency interference signals.
For a majority of the pulsars, the centre frequency of the observations was 
318 MHz; for one case it was 328 MHz, and 333 MHz for another (see Table~\ref{tbl1}).

All of the 318 MHz observations were carried out on 10-11 Dec 2000.  For most cases, the 
total number of pulses were obtained from a single observing run.  For some pulsars 
(e.g. PSR~B2045$-$16), data from different observing sessions (on different days) was combined
(after proper alignment) to obtain the final single pulse sequence with enough pulses 
for subsequent analysis.  The total number of pulses for each pulsar is given in column 7 of
Table~\ref{tbl1}.

The average profile obtained for each pulsar is shown in Figure~1.  The time resolution 
here is the basic sample time interval of 0.516 millisec, except for the case of PSR~B2111+46, 
where the data has been integrated by a factor of 4, to a sample interval of 2.064 millisec. 
The SNR obtained for the peak of the average profiles is quite good -- in the range 200 to 
600 -- resulting in single pulse peak SNRs of the order to 5 to 10 and better, which are 
needed for using the W-T technique effectively.  All the pulsars show multi-component profiles, 
with at least 3 or more components easily visible.  The classification of each pulsar, 
according to the scheme proposed by Rankin (1993), is given in col. 6 of Table~\ref{tbl1}, 
where the triple (T) pulsars have 3 known components and the multiple (M) pulsars have 5 known 
components. From published results related to (i) frequency evolution of the profile components 
and (ii) polarization properties of average profiles (e.g. Rankin 1993, Lyne \& Manchester 1988 
and references therein), it is clear for most of these pulsars that the central component is 
the core component.  Further, in almost all cases, it is apparent from Figure~1 that the conal 
components are located asymmetrically with respect to the core component, with the tendency for 
each trailing conal component to be closer to the core than its leading counterpart.  This is 
consistent with the expectations of retardation and aberration effects due to finite emission 
height of the conal beams (e.g. paper I).  

To carry out a detailed modeling of the geometry of the emission cones for each pulsar, 
we need to identify all possible emission components in the profiles, that can be 
determined within the SNR limitations of our data.
For this we have used the W-T technique described by us in paper I, where we employed this
technique to identify as many as 9 emission components in PSR~B0329+54.  
In this technique, we set a window in a desired pulse longitude range and employ an intensity 
threshold to select the single pulses that go to make an average profile, which we refer to
as a ``W-T profile''. We consider all those pulses which have emissions above the threshold 
within the window.  As a result of this averaging of selected pulses, emission components 
within the window improve in SNR compared to other parts of the profile and are thus more 
easily detected in the W-T profile.  As described in Paper I, we use a few different
checks to ensure that the detected components are genuine.  Usually, the window size and 
location are varied by a few bins while checking that the component shows up at the same
location (this is not always possible to do for components that are located in close proximity
to known, strong components).  In addition, the selection threshold is varied over a reasonable
range to check the persistence of the component, and to minimize picking up contributions from
random noise.  These checks are robust enough to avoid detection of spurious components
by chance accumulations in the selected window.  This is supported by the fact that the
W-T technique does {\it not} detect any new components when applied to off-pulse regions of 
the profile.

Using the W-T technique, we have been able to detect new emission components in most of the 
pulsars studied here, as well as get accurate estimates for the locations of the existing
components.  The total number of {\it new} components detected for each pulsar is given in 
column 8 of Table 1.  Some typical examples of detection of components (new as well as known
ones) are illustrated in the sample of W-T profiles shown in Figure~2.  Here, Figs.~2(a) \& 2(e)
show examples of clear and easy detections (usually of already known components); 
Figs.~2(b),2(d) \& 2(g) illustrate detection of new, weak components and 
Figs.~2(c), 2(f) \& 2(i) are cases of detections of new components located in close proximity 
to known, strong components.
In some cases, we were able to increase the confidence of our detections by integrating the 
data to a larger time constant (2 to 4 times the original sampling rate) and repeating the 
W-T analysis, albeit with a loss in resolution of the location of the component peak.  This 
was particularly useful in cases where the original pulse profile is relatively broad and well 
sampled (e.g. PSR~B2111+46) and the SNR is marginal for the weaker emission components.

In some cases, we detect likely candidates for new emission components but can not confirm
them with enough confidence from our data set.  This is due to limitations of either 
(i) inadequate SNR, (ii) not enough pulses for the weak components to show up reliably 
often enough or (iii) very close proximity of the suspected component to a known component, 
especially when the latter is a strong component.  In such cases, we label the component 
detection as ``tentative''.  We mention this specifically for each pulsar when we discuss 
the individual results. 

To determine accurate locations in pulse longitude for each component of a pulsar, we take
the corresponding W-T profile and fit a Gaussian curve to the selected component.  The 
noise contribution to each data point is estimated from the off-pulse noise r.m.s. of the
W-T profile.  The peak of the best-fit Gaussian is taken as the best estimate of the location 
of the component and the 1-sigma error estimate on this peak is used to derive the error in 
the location of the component.  For the sample detections shown in Figure 2, we also show 
the best fit Gaussian to the component detected within the W-T window.  For most cases, 
we get quite good fits, with values for the reduced chi-square $\sim$ 1.0.  The error 
estimates are generally found to be proportional to the SNR of the fitted component, which
is as expected.

From our determination of the total number of the emission components, we resolve the emission 
geometry for each pulsar as consisting of a central core component and one or more cones of 
emission around it.  From our best estimates of the locations of these components, we then 
solve for the emission heights of the cones, as well as for the transverse location of the 
emitting field lines on the polar cap, using the technique described in paper I.  
The polar cap location of the field lines is quantified by the ratio of the distance to 
the ``foot'' of the field line (from the magnetic pole) to the distance to the 
``foot'' of the last open field line in the magnetosphere -- the parameter \sii~ of paper I   
(in this description, the foot of the field line is the point where the field line intersects 
the neutron star surface).

\section{Results}
The results from our analysis of the 6 pulsars are summarized in Tables 2 to 7.  The table for 
each pulsar gives, for every emission cone that we detect (including the known ones), the 
following information: \phil ~\&~ \phit, the location (with respect to location of the core 
component) of the leading and trailing emission components that constitute the cone (columns 2 
and 3); \etai, the inferred retardation plus aberration angle (column 4); \gammai, the inferred 
half angle of the conal beam (column 5); \remi, the computed emission height for the cone 
(column 6); and \sii, the polar cap location of the associated field lines (column 7).  The 
emission heights of the cones are given in kilometers, as well as fraction of the light cylinder 
radius (\rlc) in column 6.  The errors for the derived quantities (columns 4 to 7) are determined 
from appropriate propagation of the errors on the estimates of the emission component locations 
(including that for the core component), and do not include any effects of possible errors in 
the values of $\alpha$ and $\beta$.

We first discuss the results for individual pulsars and present a general summary of the results
at the end of this section. 

\subsection{PSR~B0450$-$18}
This pulsar shows a 3 component profile for frequencies below about 1 GHz and is interpreted as 
a typical triple (core and one conal pair) pulsar (e.g. Rankin 1993). However, this classification
is not completely unambiguous.  Though the frequency evolution of the components (e.g. Gould, 1994)
appears to indicate that the central component is a core, the polarization characteristics do not 
show clear, unambiguous evidence of core radiation for this component.  At our frequency, the 
profile shows a fairly wide emission region around the central peak, especially on the trailing 
side (Fig.~1).  The W-T analysis readily localises the leading and trailing components in the
profile.  However, in the central 10 to 15 degrees of pulse longitude we get unclear and conflicting 
results.  The W-T analysis shows evidence for multiple emission peaks in this region (e.g. at 
$-$4.1\degsp and 3\degsp), but these are generally not found to be stable against variations of 
window widths and thresholds.  Hence, it is difficult to unambiguously identify distinct emission 
components in the central region, including a clear identification of the reference core component.
It is possible that what we are seeing in this region is a somewhat tangential cut through the 
edge of an inner emission cone, and that the core, if present at all, is a weak component in this 
region.  This would be consistent with the fact that this pulsar has a relatively large value of 
the impact parameter $\beta$ -- it is the largest amongst all the pulsars studied here (Table 1). 
The entry in column 8 of Table 1 for this pulsar reflects our uncertainty in the identification 
of new emission components.

Using a tentative identification of the core as the peak produced by setting a wide window in
the central region, we get the locations of the main leading and trailing emission components with
respect to the core as $-$8.79\degsp and +6.76\degsp, respectively.  In addition, there is some 
evidence from our analysis for the presence of weak components in the leading and trailing wings 
of the profile, but these can not be confirmed with the quality of our data.  Solving for a single 
emission cone for this pulsar, the emission height and polar cap location are obtained as 310 km 
(1.2\% of \rlc) and 0.56 respectively (see Table~\ref{tbl_0450}).  We note that the results for 
this pulsar should be treated with caution, as the emission geometry is probably more complicated 
than what we present. 

\subsection{PSR~B1237+25}
The typical average profile for this well studied pulsar clearly shows 5 emission components 
with the central one known to be the core component.  The centers of both cones are clearly 
offset to earlier longitudes than the core, as would be expected due to retardation and 
aberration.  Our analysis reveals the presence of two more emission components, located between 
the core and the inner conal ring.  Of these two, the component on the leading side 
(at $-$2.12\degsp) is seen relatively easily in our analysis (Fig.~2b), whereas the one on the 
trailing side (at +1.56\degsp) is somewhat harder to detect.  This is because it is located in 
the relatively narrower region between the core and the trailing component of the inner cone.  
We therefore interpret this pulsar as having 3 emission cones around the core component, 
with each cone clearly showing retardation and aberration effects.
Solving for the locations of the emission cones at 318 MHz gives heights of 180, 460 and 
600 km (0.3\%, 0.7\% and 0.9\% of \rlc ~ respectively) and transverse locations on the 
polar cap of 0.33, 0.41 and 0.59 (see Table~\ref{tbl_1237}).

\subsection{PSR~B1821+05}
The average profile for this pulsar shows a core dominated triple component emission beam 
at 318 MHz (Fig.~1).  The conal components, located at $-$13.75\degsp and +10.94\degsp,
clearly exhibit retardation-aberration effects. Even though the SNR is not very good, there 
are indications of extra emission components between the core and the known conal components.
With some difficulty, our analysis picks up a component at $-$6.07\degsp from the core and 
one at +8.2\degsp on the trailing side (Figs.~2c \& 2d).  In addition, we have marginal 
detections of two more conal components, at $-$10.25\degsp and +4.61\degsp.  These last two 
are detected with fairly low thresholds in the W-T technique and are, at best, tentative 
detections, requiring confirmation with higher sensitivity observations.  Analysis of the 
data after further integration by a factor of 2 were used to confirm the presence of these 
components.

Our final result for this pulsar is a detection of 3 cones for which the emission heights 
at 318 MHz are obtained as 290, 410 and 570 km (0.8\%, 1.2\% and 1.6\% of \rlc ~ respectively), 
and polar cap locations are obtained as 0.43, 0.57 and 0.63 respectively 
(see Table~\ref{tbl_1821}).

\subsection{PSR~B1857$-$26}
This is a well known multi-component (M-type) pulsar, where the typical profile at meter 
wavelengths readily shows 4 components and has clear indications of a fifth component located 
close to the leading edge of the central, core component (see Fig.~2e for our detection of 
this component).  Our analysis easily gives the locations for these 5 known components.
In addition, we see evidence for two more outer components, located near the leading and 
trailing edges of the profile, at $-$23.73\degsp and +19.95\degsp with respect to the core 
(see Fig.~2f for one of these).  We label these two as tentative detections, as the threshold 
and number of contributing pulses are rather small (though the analysis of data integrated by 
a factor of 2 supports these detections).  Further, there are hints of emission components 
in between the known conal components, but these can not be confirmed with the quality of 
our data.

We thus have 5 confirmed components (along with 2 tentative ones) for this pulsar, leading
to an interpretation of core and 2 emission cones (and a tentative third outer cone).  
The emission heights (at 318 MHz) for these 3 cones are found to be 220, 480 and 690 km 
(0.8, 1.7 and 2.4 \% of \rlc) and polar cap locations of 0.61, 0.69 and 0.74 (Table~\ref{tbl_1857}).

\subsection{PSR~B2045$-$16}
This well known pulsar shows a 3 component profile where the central core component is
clearly asymmetrically located with respect to the centre of the conal ring.  This is 
an indication of significant aberration and retardation effects for the emission cone.
On using the W-T technique, we are able to detect 3 more emission components, located in 
the ``saddle'' regions between the core and the known conal components, at longitudes of
$-$7.2\degsp, $-$5.5\degsp and +3.06\degsp with respect to the core (see Fig.~2g 
for an example of one of these).  This gives a total of 5 conal 
components, leaving us with the somewhat difficult situation of a probable undetected conal 
component, which needs to be located in the narrow valley between the core and the already
known trailing conal component. There are some indications of an emission component at 
+4.0\degsp, but we can not determine this unambiguously from our analysis, as it is too close
to the known component at +4.59\degsp.  Lacking this, we have the problem of deciding which 
of the 3 new conal components is the odd one out.  If we take the components at 
$-$5.5\degsp and +3.06\degsp as forming the second conal ring, the solutions for the two rings 
are: emission heights at 1240 and 2230 km (1.3 and 2.4 \% of \rlc) and polar cap locations 
of 0.28 and 0.32 -- this is the solution listed in Table~\ref{tbl_2045}.  
Alternatively, taking the components at $-$7.2\degsp and +3.06\degsp as forming the second conal 
ring changes the emission height estimate for this cone from 1240 km to 2090 km and the polar 
cap location shifts from 0.28 to 0.22.  We believe the first option is more likely to be the 
real situation.  

\subsection{PSR~B2111+46}
This pulsar shows a very wide but clearly triple profile with a dominant core component at meter 
wavelengths (Fig.~1).  The location of the conal components is clearly asymmetric with respect 
to the core, showing retardation-aberration effects.  For this pulsar two data sets, one taken 
at 318 MHz in December 2000 and another taken at 333 MHz in February 2002 were available.  Since 
the data from February 2002 had better SNR and more number of pulses, all the results presented 
here are from the analysis of this data set.  Analysis of the December 2000 data produced results
which are similar to those reported here, and well within the errors of the measurements.

Since the profile is very wide, the data were averaged to 4 times the original time resolution 
of 0.516 millisec, to improve the SNR for the W-T analysis.  Besides confirming the location of
the main components, our W-T analysis picks up 2 additional conal components, located at 
$-$18.71\degsp and +14.91\degsp with respect to the core (see Figs.~2h \& 2i).  In addition, 
there are hints of another pair of components located at the leading and trailing edges of the 
profile (at approximately $-$48.3\degsp and +38.7\degsp with respect to the core), but we are 
unable to confirm them clearly.  Hence we confine ourselves to 2 cones of emission around the 
central core component, for this pulsar.  The estimates for the emission heights at 333 MHz are 
1360 km and 2080 km (2.8 and 4.3 \% of \rlc), while the values for the polar cap locations are 
0.22 and 0.31 (see Table~\ref{tbl_2111}).

\subsection{Summary of results}
Our main results can be summarized as follows:

(i) All the 6 pulsars show clear signatures of an asymmetry of the cones with respect to 
the core radiation, in that the centers of the cones occur at earlier pulse longitudes, 
with respect to the location of the core. For a given pulsar, the magnitude of the effect 
increases from inner to outer cones.

(ii) Interpreting the shift of the cone centers as the effect of retardation and 
aberration on the emitted conal beam leads to reasonable values for the emission 
heights.  These range from about 200 km to 2200 km over all the 6 pulsars.  In terms
of the light cylinder distance, this range is from 0.3\% to 4.3\% of \rlc. For a 
given pulsar, the heights are found to increase systematically from the inner to the 
outer cones (though in some cases the sizes of the error bars prevent this from being 
made a firm conclusion).

(iii) Combining the emission height estimates with a dipole field line geometry 
results in estimates of the transverse polar cap locations of the foot of the field 
lines that each cone is radiated from.  These are found to vary from about 0.22 to 
0.74 of the radial distance to the last open field line from the magnetic axis.  
Furthermore, this value is found to increase systematically from inner to outer 
cones for a given pulsar (though again, for some cases, the sizes of the errors 
reduce the significance of this conclusion).

\begin{table}
\begin{center}
\caption{Emission geometry results for PSR~B0450$-$18
\label{tbl_0450}}
\begin{tabular}{crrrrrr}
\tableline\tableline
Cone &\multicolumn{1}{c}{\phil} &\multicolumn{1}{c} {\phit} &\multicolumn{1}{c}{\etai} &\multicolumn{1}{c} {\gammai} &
\multicolumn{1}{c}{\remi}   & \multicolumn{1}{c}{\sii} \\
No.  &\multicolumn{1}{c}{(deg)} &\multicolumn{1}{c} {(deg)} &\multicolumn{1}{c}{(deg)} &\multicolumn{1}{c}{(deg)}  &
\multicolumn{1}{c}{~~(km) ~~(\% \rlc)} &                   \\
\tableline
1 & $-$8.79 $\pm$ 0.10 & 6.76 $\pm$ 0.11 & $-$1.01 $\pm$ 0.11 & 5.25 $\pm$ 0.02 & 310 $\pm$ 30 (1.2\%)& 0.56 $\pm$ 0.03 \\
\tableline
\end{tabular}
\end{center}
\end{table}

\begin{table}
\begin{center}
\caption{Emission geometry results for PSR~B1237+25
\label{tbl_1237}}
\begin{tabular}{crrrrrr}
\tableline\tableline
Cone &\multicolumn{1}{c}{\phil} &\multicolumn{1}{c} {\phit} &\multicolumn{1}{c}{\etai} &\multicolumn{1}{c} {\gammai} &
\multicolumn{1}{c}{\remi}   & \multicolumn{1}{c}{\sii} \\
No.  &\multicolumn{1}{c}{(deg)} &\multicolumn{1}{c} {(deg)} &\multicolumn{1}{c}{(deg)} &\multicolumn{1}{c}{(deg)}  &
\multicolumn{1}{c}{~~(km) ~~(\% \rlc)} &                   \\
\tableline
1 & $-$2.12 $\pm$ 0.07 & 1.56 $\pm$ 0.09 & $-$0.28 $\pm$ 0.07 & 1.47 $\pm$ 0.05 & 180 $\pm$ 40 (0.3\%)& 0.33 $\pm$ 0.04 \\
2 & $-$4.41 $\pm$ 0.03 & 2.98 $\pm$ 0.06 & $-$0.72 $\pm$ 0.05 & 2.95 $\pm$ 0.03 & 460 $\pm$ 30 (0.7\%)& 0.41 $\pm$ 0.02 \\
3 & $-$6.97 $\pm$ 0.02 & 5.11 $\pm$ 0.02 & $-$0.93 $\pm$ 0.04 & 4.82 $\pm$ 0.01 & 600 $\pm$ 30 (0.9\%)& 0.59 $\pm$ 0.01 \\
\tableline
\end{tabular}
\end{center}
\end{table}

\begin{table}
\begin{center}
\caption{Emission geometry results for PSR~B1821+05
\label{tbl_1821}}
\begin{tabular}{crrrrrr}
\tableline\tableline
Cone &\multicolumn{1}{c}{\phil} &\multicolumn{1}{c} {\phit} &\multicolumn{1}{c}{\etai} &\multicolumn{1}{c} {\gammai} &
\multicolumn{1}{c}{\remi}   & \multicolumn{1}{c}{\sii} \\
No.  &\multicolumn{1}{c}{(deg)} &\multicolumn{1}{c} {(deg)} &\multicolumn{1}{c}{(deg)} &\multicolumn{1}{c}{(deg)}  &
\multicolumn{1}{c}{~~(km) ~~(\% \rlc)}  &                   \\
\tableline
1 &  $-$6.07 $\pm$ 0.47 &  4.61 $\pm$ 0.30 & $-$0.73 $\pm$ 0.31 & 3.36 $\pm$ 0.13 & 290 $\pm$ 120 (0.8\%) & 0.43 $\pm$ 0.09 \\
2 & $-$10.25 $\pm$ 0.30 &  8.20 $\pm$ 0.43 & $-$1.03 $\pm$ 0.29 & 5.28 $\pm$ 0.13 & 410 $\pm$ 120 (1.2\%) & 0.57 $\pm$ 0.08 \\
3 & $-$13.75 $\pm$ 0.39 & 10.94 $\pm$ 0.27 & $-$1.41 $\pm$ 0.27 & 6.90 $\pm$ 0.12 & 570 $\pm$ 110 (1.6\%) & 0.64 $\pm$ 0.06 \\
\tableline
\end{tabular}
\end{center}
\end{table}

\begin{table}
\begin{center}
\caption{Emission geometry results for PSR~B1857$-$26
\label{tbl_1857}}
\begin{tabular}{crrrrrr}
\tableline\tableline
Cone &\multicolumn{1}{c}{\phil} &\multicolumn{1}{c} {\phit} &\multicolumn{1}{c}{\etai} &\multicolumn{1}{c} {\gammai} &
\multicolumn{1}{c}{\remi}   & \multicolumn{1}{c}{\sii} \\
No.  &\multicolumn{1}{c}{(deg)} &\multicolumn{1}{c} {(deg)} &\multicolumn{1}{c}{(deg)} &\multicolumn{1}{c}{(deg)}  &
\multicolumn{1}{c}{~~(km) ~~(\% \rlc)}  &                   \\
\tableline
1 &  $-$9.78 $\pm$ 0.25 &  8.51 $\pm$ 0.25 & $-$0.64 $\pm$ 0.20 &  4.58 $\pm$ 0.07 & 220 $\pm$  70 (0.8\%) & 0.61 $\pm$ 0.10 \\
2 & $-$18.06 $\pm$ 0.11 & 15.30 $\pm$ 0.07 & $-$1.38 $\pm$ 0.11 &  7.64 $\pm$ 0.03 & 480 $\pm$  40 (1.7\%) & 0.69 $\pm$ 0.03 \\
3 & $-$23.73 $\pm$ 0.46 & 19.95 $\pm$ 0.34 & $-$1.96 $\pm$ 0.30 &  9.77 $\pm$ 0.12 & 690 $\pm$ 100 (2.4\%) & 0.74 $\pm$ 0.06 \\
\tableline
\end{tabular}
\end{center}
\end{table}

\begin{table}
\begin{center}
\caption{Emission geometry results for PSR~B2045$-$16
\label{tbl_2045}}
\begin{tabular}{crrrrrr}
\tableline\tableline
Cone &\multicolumn{1}{c}{\phil} &\multicolumn{1}{c} {\phit} &\multicolumn{1}{c}{\etai} &\multicolumn{1}{c} {\gammai} &
\multicolumn{1}{c}{\remi}   & \multicolumn{1}{c}{\sii} \\
No.  &\multicolumn{1}{c}{(deg)} &\multicolumn{1}{c} {(deg)} &\multicolumn{1}{c}{(deg)} &\multicolumn{1}{c}{(deg)}  &
\multicolumn{1}{c}{~~(km) ~~(\% \rlc)}  &                   \\
\tableline
1 & $-$5.50 $\pm$ 0.28 & 3.06 $\pm$ 0.13 & $-$1.22 $\pm$ 0.16 & 2.78 $\pm$ 0.08 & 1240 $\pm$ 160 (1.3\%) & 0.28 $\pm$ 0.02 \\
2 & $-$8.97 $\pm$ 0.04 & 4.59 $\pm$ 0.03 & $-$2.19 $\pm$ 0.05 & 4.18 $\pm$ 0.01 & 2230 $\pm$  50 (2.4\%) & 0.32 $\pm$ 0.004 \\
\tableline
\end{tabular}
\end{center}
\end{table}

\begin{table}
\begin{center}
\caption{Emission geometry results for PSR~B2111+46
\label{tbl_2111}}
\begin{tabular}{crrrrrr}
\tableline\tableline
Cone &\multicolumn{1}{c}{\phil} &\multicolumn{1}{c} {\phit} &\multicolumn{1}{c}{\etai} &\multicolumn{1}{c} {\gammai} &
\multicolumn{1}{c}{\remi}   & \multicolumn{1}{c}{\sii} \\
No.  &\multicolumn{1}{c}{(deg)} &\multicolumn{1}{c} {(deg)} &\multicolumn{1}{c}{(deg)} &\multicolumn{1}{c}{(deg)}  &
\multicolumn{1}{c}{~~(km) ~~(\% \rlc)}  &                   \\
\tableline
1 & $-$18.71 $\pm$ 0.72 & 14.91 $\pm$ 0.53 & $-$1.90 $\pm$ 0.45 & 3.14 $\pm$ 0.07 & 1360 $\pm$ 320 (2.8\%) & 0.22 $\pm$ 0.03 \\
2 & $-$35.69 $\pm$ 0.11 & 29.87 $\pm$ 0.16 & $-$2.91 $\pm$ 0.11 & 5.61 $\pm$ 0.02 & 2080 $\pm$ 80 (4.3\%) & 0.31 $\pm$ 0.006 \\
\tableline
\end{tabular}
\end{center}
\end{table}

\section{Discussion}
There are several interesting aspects of pulsar emission geometry that can be addressed 
by our study.  Firstly, it should be clear from this work (and paper I) that it is not 
sufficient to study only the average profiles to understand the emission geometry.  
We have shown that a careful analysis of good quality single pulse data can provide 
valuable extra information about the emission geometry.  It would be almost impossible 
to detect some of the emission components we have detected, using other techniques such 
as fitting multiple gaussians to the average profile.  Here, the window-thresholding 
method used by us comes in as a handy tool to find components that are weak and/or emit 
intermittently, or those which are located close to other strong components.  Using this, 
we find that most triple and multiple component pulsars can be shown to have 5 or 7 
components.  It is then likely that most pulsars, when observed with sufficient quality 
data, are going to show multiple emission cones around the central core component.  

Further, a careful identification of the emission components appears, almost invariably, to 
support the picture of a core beam plus multiple, concentric conal beams of emission, in
that the conal components turn out to be arranged in equal numbers on either side of the
core.  What is even more interesting is that in all cases, there is an asymmetry in the
location of the core with respect to the centers of the cones, and that too always in the
same sense of having the cone centers advanced with respect to the core component location.
Though we have presented the detailed analysis for only a handful of pulsars, this trend
is easily visible in a vast majority of pulsars that have clearly defined core and conal 
emission components.  A check of the data from different sources in the literature 
(e.g. \cite {G94}; \cite {HR}; \cite {Weisberg}; the EPN database) reveals several clear 
cases: PSRs~B0105+68,  B1700$-$32, B1737+13, B1826$-$17, B1916+14, B1946+35, B2002+31 \& 
B2003$-$08 (detailed studies of several such pulsars have been initiated and will be reported 
upon shortly).  Thus it would seem that this is a fairly universal trend among pulsars that
show clear core and conal emission components, and our interpretation of this as being due
to retardation-aberration effects in the magnetosphere, is a reasonable one.

The emission heights that we obtain for the different pulsars have very plausible values.  
For PSRs~B0450$-$18, B1237+35, B1821+05 and B1857$-$26, we obtain emission heights in the range 
of 200 to 700 km, while for PSRs~B2045$-$16 and B2111+46, the values range from 1200 to 2200 km.  
Typical estimates of emission heights reported in literature are a few 100 km for most 
pulsars (e.g. \cite{JMRd},  \cite{KG2}).  
Rankin (1993) estimates the emission heights for the cones from the measured half-power 
widths of the conal component pairs and making the assumption that the cones are radiated 
from near the last open field lines of the polar cap.  This yields 1 GHz 
height estimates of about 130 and 220 km for the two cones.  
Our height estimates will naturally be larger than these as we find that the emission 
is located on field lines that are well inside of the last open field line: 
a cone of the same angular width radiated from inner field lines has to come from 
a greater height in the magnetosphere.  
Specifically, for PSRs~B1237+25 and B1857$-$26, Rankin(1993) gives estimates of emission 
heights for the cones at a range of frequencies.  Comparing with the values at the 
nearby frequency of 270 MHz (Table 3 of Rankin, 1993), we find that our estimates for 
the same cone at 318 MHz are more than twice as large.  If, however, the estimates 
of Rankin (1993) are corrected for emitting field line with $\sii ~ \sim ~ 0.6$
(as postulated in \cite{MR02}), then the results are in better agreement
with ours.
Our emission height estimates are closer to those obtained by Kijak \& Gil (1997,1998), 
though still somewhat larger.  

We note that the height estimates for PSRs~B2045$-$16 and B2111+46 are somewhat unusually large.
Correspondingly, these two pulsars also have rather small values for the polar cap locations 
of the radiating field lines, as compared to the lower emission height pulsars.  The reason 
for these larger heights is not clear.
As a fraction of the light cylinder radius distance, however, all these heights are still
quite small, being $\sim$ 5\% of \rlc~ at the largest.  This also means that effects of magnetic
field sweepback (see paper I) can safely be ignored for all these cases.

Strictly speaking, all our height estimates are heights of the cones {\it relative to that 
of the core radiation}, as the retardation-aberration of the cone centers is measured 
with the core component as reference.  Since the core radiation is thought to
originate very close to the neutron star surface (e.g. Rankin, 1993 \& 1990), these are
good estimates for the actual emission heights of the cones above the neutron star surface
(the extra height of 10 km for the radius of the neutron star is much smaller than our
typical error bars).
However, if this assumption about the height of the core emission beam is not 
correct, then the height of the core emission has to be added to the height estimates we 
have reported for the cones  -- in this sense, our results can be taken as lower limits to 
the actual emission heights.  The other effect of a finite core height will be to cause the 
inferred polar location of the cones to become somewhat smaller (see Eqn. 15 of paper I), 
with the fractional change diminishing progressively from inner to outer cones.  For example, 
an emission height of 100 km for the core radiation of PSR~B1237+25 would cause the polar 
cap locations of the cones to reduce to 0.26, 0.37 and 0.55 from their present values of 
0.33, 0.41 and 0.59.
In any case, it is quite clear from our results that the core emission originates at heights 
that are substantially ($\sim$ 100s of km) less than that for the cones -- otherwise we 
would not see any relative retardation-aberration effects between the core and cones.

Another interesting feature of the results is that they are quite self-consistent 
with radio emission originating within the open field line region of the magnetosphere, 
in that we do not obtain results for polar cap locations that are greater than 1.0.  This 
can happen in principle, as the emission height estimate is completely independent of 
the width of the cone, whereas both these parameters are relevant for the calculation of 
the polar cap location (see Eqns. 9 and 15 of paper I).
Polar cap locations greater than 1.0 can result if the emission height estimates are
less than the current results by factors of 4 to 16.  In a sense then, this argues that 
our interpretation of component location asymmetries as being due to retardation-aberration 
effects is a sensible interpretation.

Perhaps the most interesting result from this work is that the different cones for a pulsar
originate on different sets of field lines in the magnetosphere.  Clearly, all the conal 
radiation does {\it not} originate near the last open field line region.  In fact, it appears
that a significant fraction of the polar cap region may be active in the generation of the
multiple conal beams.  Furthermore, our results indicate that this activity is confined 
along distinct and concentric rings on the polar cap.  Generally, it is thought
(e.g. \cite {RS75}, \cite {CR80}) that this activity may be produced by spark discharges
in vacuum gaps formed immediately above the neutron star surface.  
For such a model, our results indicate that the sparks are confined to concentric 
rings on the polar cap.
We note that there are recent reports of inference of such patterns of circulating 
sparks for pulsars with conal-single geometries showing drifting sub-pulses 
(e.g. \cite {DR99}).  Our results tend to support a similar conclusion
for multi-component pulsars with more central traverses of the observing line-of-sight.
For our pulsars, we can actually constrain the radii of these rings, via the \sii~ parameter.
For a 10 km radius neutron star with a period of P seconds, the radial size of the polar cap 
is just $\rm 145 ~P^{-0.5}$ meters.   For three concentric conal rings anchored to field 
lines at \sii~ = 0.33, 0.41 and 0.59 (as in PSR~B1237+25), this translates to ring radii 
of 41, 51 and 73 meters.  Similar numbers ($\sim$ a few to several tens of meters) are 
obtained for other pulsars also.  It is indeed remarkable that we can infer the presence 
of features of this size on the polar cap.

\cite{GS2000} have recently proposed a model for pulsar radiation that invokes 
concentric rings of sparks to explain the observed properties of pulsar radiation. 
In their theory, the separation between concentric rings of sparks, very close to 
the neutron star surface, is of the order of {\it h}, the height of the vacuum gap.
They estimate the value of {\it h} to be of the order of 10 meters for pulsars with 
period $\sim$ 1 second, which is quite similar to the typical separation of rings 
that we infer.  For example, for PSR~B1237+25, the separation between first and second
rings is 10 meters and that between the second and third is 22 meters.  For PSR~B1821+05,
these numbers are 23 and 12 meters respectively; for PSR~B1857$-$26, they are 15 and 9
meters.  Thus our results make interesting connections with some of the existing 
theoretical models and support the idea of concentric rings of sparks circulating
on the polar cap as being the cause of the radio radiation.

Although the activity that is the ultimate source of the radio radiation that we 
observe may be occurring very close to the neutron star surface, the final radiation that 
we see appears to originate at significant heights in the magnetosphere.  Further, we 
have shown that the inner cones are emitted at lower altitudes and outer cones at 
progressively higher altitudes. In addition, it is generally accepted that the same cone 
is emitted at higher heights at lower frequencies.  Clearly, all this is pointing to the 
requirement of the appropriate combination of parameters at specific points in the 
magnetosphere for the plasma disturbance to produce a beam of radiation that can be seen 
along our line of sight through the magnetosphere.  This could be a combination of physical 
and geometrical parameters that contrive to give the right conditions at specific points 
in the magnetosphere.  The localisation of the emitting regions achieved by the kind
of work reported here and in paper I should help in better constraining the parameter 
space and thus help the theories trying to find the correct explanation for the generation
of the radio waves.

\section{Conclusions}
Using the window-thresholding technique to analyse single pulse data from observations in the
325 MHz band of the GMRT, we have conducted a detailed analysis of the emission geometry of 6 
radio pulsars that have prominent, multiple-component profiles.  We have been able to detect new
emission components in almost all of the pulsars.  In some cases, we report tentative detections
of weak components that need further confirmation studies.  It thus looks likely that many 
multiple component pulsars may have at least 5 to 7 components. The total number and location of 
the emission components we detect clearly supports a picture of multiple conal beams around a 
central core beam for these pulsars (and are a pointer to the possibility that most pulsars
may have such an emission beam structure).  

Further, for all cases of conal emission, we find that the cone centre is offset to earlier
longitudes with respect to the location of the core, and this effect increases in magnitude
from the innermost to the outermost cones for every pulsar.  It appears this effect may be
present in most pulsars that have clear core and conal emission geometries. We interpret this 
effect as being due to retardation and aberration of the conal emission beams in the magnetosphere 
of the pulsar.  From this, we are able to solve for the emission geometry and estimate the emission 
height for each cone and also the polar cap location of the field lines associated with the cone,
following the technique developed in Gangadhara \& Gupta (2001).
We find emission height estimates in the range of 200 to 700 km for four pulsars -- PSRs~B0450$-$18, 
B1237+25, B1821+05 and B1857$-$26 -- and 1200 to 2200 km for two others -- PSRs~B2045$-$16 
and B2111+46.  In terms of the light cylinder distance, the emission altitudes range from 0.3\% 
to 4.3\%.  The estimates for the polar cap location of the radiating field lines lines come out 
to be in the range 0.22 to 0.74 of the distance to the last open field line, from the magnetic 
axis.  Further, for every pulsar with multiple cones, the emission altitude and the polar cap 
location are found to increase systematically from the innermost to the outermost cone.

Mapping the locations of the field lines associated with each cone, on to the neutron surface, 
gives concentric rings with radii of the order of a few to several tens of meters and spacings 
between adjacent rings of about 10 to 20 meters.  Our results thus tend to support a model of 
concentric rings of sparks produced in the vacuum gap region just above the neutron surface 
(e.g. \cite{GS2000}).

The fairly precise localization of emission regions that we have reported should provide 
additional information for constraining the parameter space for theories and models of 
radio emission mechanisms.

\begin{acknowledgements}
We thank P. Ahmadi and the staff of the GMRT for help with the observations. 
The GMRT is run by the National Centre for Radio Astrophysics of the Tata Institute 
of Fundamental Research.  We wish to thank the referee for useful suggestions that 
helped improve the quality of the results, and J. Gil and T. Hankins for comments on the 
manuscript.  In addition, YG would like to acknowledge the support of the Arecibo Observatory 
of the NAIC for a sabbatical visit, during which the bulk of the work on this manuscript 
was done.
\end{acknowledgements}

\clearpage

\begin{figure}
\plotone{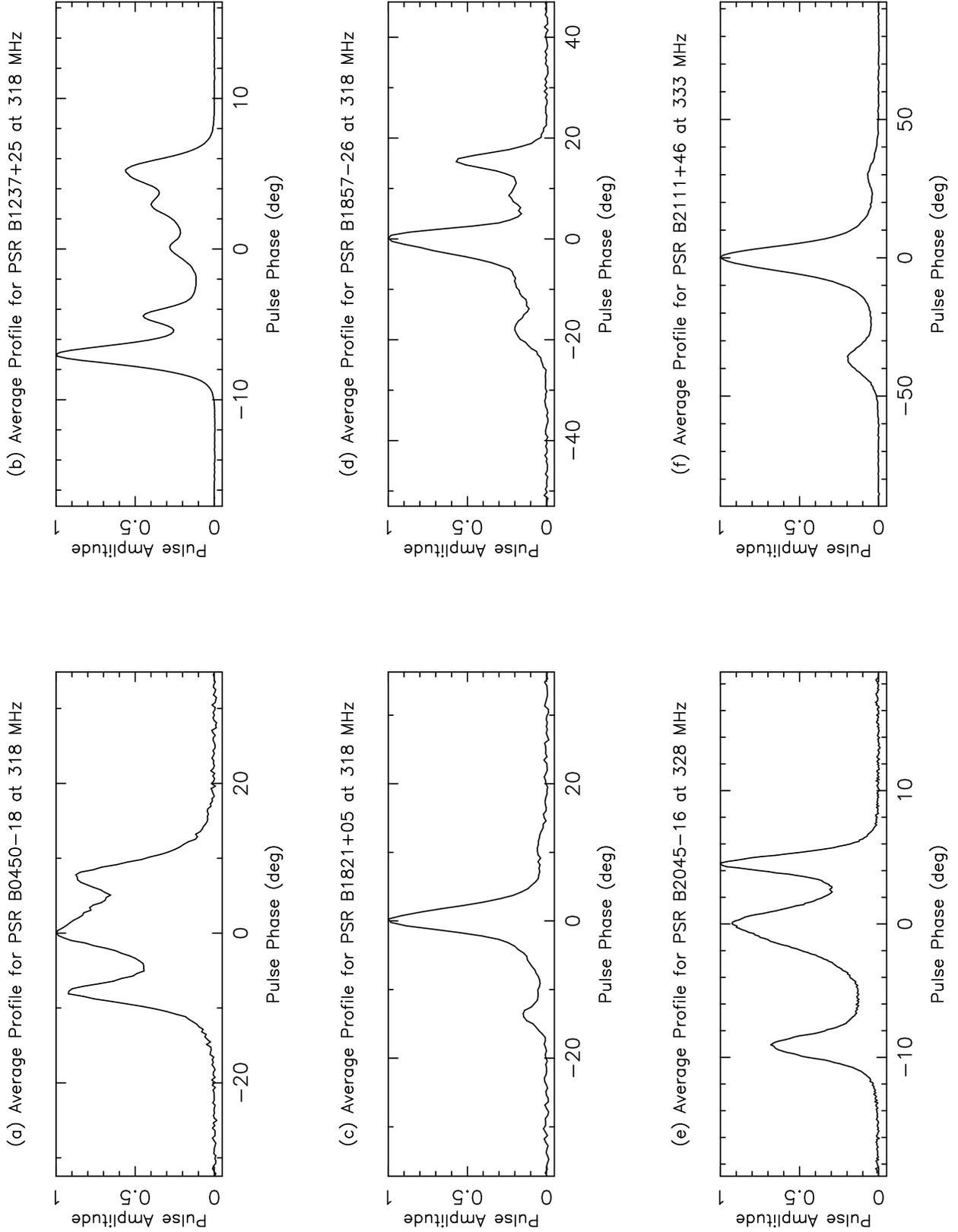}
\caption{Average pulse profiles for the 6 pulsars. The time resolution is 0.516 millisec for all
except PSR~B2111+46, for which it is 2.064 millisec. \label{fig1}}
\end{figure}

\begin{figure}
\plotone{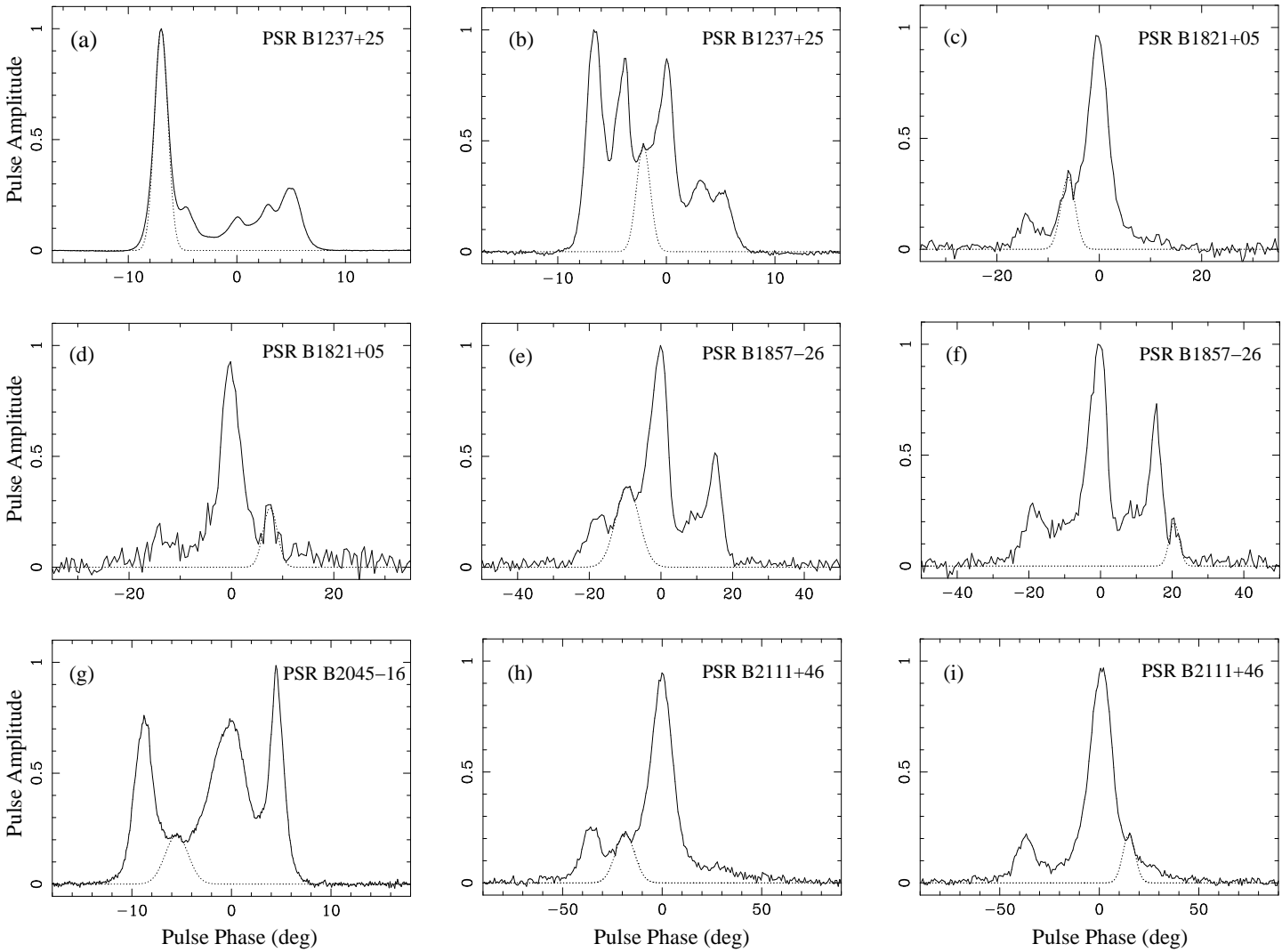}
\caption{Examples of detection of emission components.  The solid curves are the W-T profiles and
the dotted curves show the best fit Gaussian model for the detected component.  \label{fig2}}
\end{figure}
\end{document}